\documentclass[aps,prd,reprint,superscriptaddress,nofootinbib,amsfonts,amssymb,amsmath,twocolumn]{revtex4-1}

\usepackage[utf8x]{inputenc}
\usepackage[dvipdf,dvips,dvipdfmx]{graphicx}
\usepackage{float}
\usepackage{wrapfig}
\usepackage{bm}
\usepackage{color}
\usepackage{comment}
\usepackage{xcolor}
\usepackage{hyperref}
\hypersetup{
colorlinks=true,
citecolor=blue,
citebordercolor=red,
linktoc=all,
linkcolor=blue,
urlcolor=blue
}

\catcode`\@=11
\def\lsim{\mathrel{\mathpalette\@versim<}}
\def\gsim{\mathrel{\mathpalette\@versim>}}
\def\@versim#1#2{\vcenter{\offinterlineskip
\ialign{$\m@th#1\hfil##\hfil$\crcr#2\crcr\sim\crcr } }}
\catcode`\@=12
\newcommand{\Slash}[1]{{\ooalign{\hfil/\hfil\crcr$#1$}}}

\newcommand{\p}{\partial}

\newcommand{\bp}{\begin{pmatrix}}
\newcommand{\ep}{\end{pmatrix}}
\newcommand{\nn}{\nonumber\\}

\newcommand{\df}{\text{d}}

\newcommand{\bs}[1]{\boldsymbol}

\newcommand{\pmat}[1]{\begin{pmatrix}#1\end{pmatrix}}

\newcommand{\be}{\begin{equation}}
\newcommand{\ee}{\end{equation}}
\newcommand{\bea}{\begin{eqnarray}}
\newcommand{\eea}{\end{eqnarray}}

\newbox{\ORCIDicon}
\sbox{\ORCIDicon}{\large
                  \includegraphics[width=0.8em]{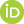}}

\begin{document}

\title{
Neutrino mass generation in asymptotically safe gravity
}
\author{Gustavo P. de Brito\,\href{https://orcid.org/0000-0003-2240-528X}{\usebox{\ORCIDicon}}}
\email{gp.brito@unesp.br}
\affiliation{Departamento de F\'isica, Universidade Estadual Paulista (Unesp), Campus Guaratinguet\'a,
Av.~Dr.~Ariberto Pereira da Cunha, 333, Guaratinguet\'a, SP, Brazil.}
\author{Astrid Eichhorn\,\href{https://orcid.org/0000-0003-4458-1495}{\usebox{\ORCIDicon}}}
\email{eichhorn@thphys.uni-heidelberg.de}
\affiliation{Institut f{\"u}r Theoretische Physik, Universit{\"a}t Heidelberg, Philosophenweg 16, 69120 Heidelberg, Germany}
\author{Antonio D. \surname{Pereira}\,\href{https://orcid.org/0000-0002-6952-2961}{\usebox{\ORCIDicon}}}
\email{adpjunior@id.uff.br}
\affiliation{Instituto de F\'isica, Universidade Federal Fluminense, Campus da Praia Vermelha, Av. Litor\^anea s/n, 24210-346, Niter\'oi, RJ, Brazil}
\author{Masatoshi \surname{Yamada}\,\href{https://orcid.org/0000-0002-1013-8631}{\usebox{\ORCIDicon}}\,}
\email{m.yamada@kwansei.ac.jp}
\affiliation{Department of Physics and Astronomy, Kwansei Gakuin University, Sanda, Hyogo, 669-1330, Japan}

\begin{abstract} 
There exist several distinct phenomenological models to generate neutrino masses.
We explore, which of these models can consistently be embedded in a quantum theory of gravity and matter. 
We proceed by invoking a minimal number of degrees of freedom beyond the Standard Model.
Thus, we first investigate whether the Weinberg operator, a dimension-five-operator that generates neutrino masses without requiring degrees of freedom beyond the Standard Model, can arise in asymptotically safe quantum gravity. We find a negative answer with far-reaching consequences: 
new degrees of freedom beyond gravity and the Standard Model are necessary to give neutrinos a mass in the asymptotic-safety paradigm. Second, we explore whether the type-I Seesaw mechanism is viable and discover an upper bound on the Seesaw scale. 
The bound depends on the mass of the visible neutrino.
We find a numerical value of $10^{14}\, \rm GeV$ for this bound when neglecting neutrino mixing for a visible mass of $10^{-10}\, \rm GeV$.
Conversely, for the most ``natural" value of the Seesaw scale in a quantum-gravity setting, which is the Planck scale, we predict an upper bound for the neutrino mass of the visible neutrino of approximately $10^{-15}\, \rm GeV$.
Third, we explore whether neutrinos could also be Pseudo-Dirac-neutrinos in asymptotic safety and find that this possibility can be accommodated. 

\end{abstract}
\maketitle

\section{Introduction}
Neutrinos have masses. 
Their differences are measured with neutrino oscillations~\cite{K2K:2002icj,KamLAND:2004mhv,MINOS:2008kxu,MINOS:2020llm,RENO:2012mkc,DoubleChooz:2012gmf,DayaBay:2012fng,T2K:2013ppw,Super-Kamiokande:1998kpq,SNO:2002tuh,OPERA:2018nar,NOvA:2021nfi,Capozzi:2016rtj,IceCubeCollaboration:2023wtb}. An upper bound on the sum of the three non-sterile neutrino masses is obtained from cosmological surveys~\cite{Planck:2018vyg,DESI:2024mwx}. 
Recently, the Dark Energy Spectroscopic Instrument (DESI) survey~\cite{DESI:2025ejh} has reduced this upper bound to $0.064 \, \rm eV$ (under the assumption of $\Lambda$CDM), close to the lower bound of $0.059 \, \rm eV$ in the normal hierarchy \cite{Esteban:2024eli} and $0.10 \, \rm eV$ in the inverted hierarchy \cite{deSalas:2020pgw}. The KATRIN experiment provides an upper bound of 0.45 eV for the effective electron neutrino mass from beta decay~\cite{Katrin:2024tvg,KATRIN:2021uub}.
For comprehensive reports on massive neutrinos, see \cite{Mohapatra:1991ng,Fukugita:2003en,Mohapatra:2004ht,Mohapatra:2005wg,Mohapatra:2006gs,Gonzalez-Garcia:2007dlo,Abazajian:2012ys,Drewes:2013gca,Xing:2014wwa}.

The origin of neutrino masses is unknown. The ratio of neutrino masses to the other Standard Model (SM) fermion masses is tiny~\cite{ParticleDataGroup:2022pth}.  Electroweak symmetry breaking that endows neutrinos with a standard Dirac mass term has long been considered unappealing, because no mechanism was known to explain this tiny ratio. Recently, a mechanism was proposed in quantum gravity:
In an asymptotically safe theory of gravity and matter, neutrino Yukawa couplings may dynamically be driven towards tiny values~\cite{Held:2019vmi,Kowalska:2022ypk,Eichhorn:2022vgp}. 
Another hint that neutrino masses may be connected to quantum gravity comes from the Seesaw mechanism, where the mass scale of right-handed neutrinos is typically assumed to be $\mathcal{O}(10^{16}\, \mathrm{GeV})$ and thus close to the Planck scale.

 We take these results as motivation to explore the relation between quantum gravity and neutrino mass generation in more depth. Our starting point is that in asymptotically safe theories, the couplings are constrained at the Planck scale. The Renormalization Group (RG) flow translates these constraints to low energy scales, where testable consequences for particle physics arise.

\section{Background on neutrino masses in asymptotically safe theories with gravity}
Asymptotically safe theories rely on an interacting fixed point of the RG flow of gravity and matter at trans-Planckian scales. There is a vast body of work providing convincing evidence for its existence, see \cite{Eichhorn:2017egq,Eichhorn:2018yfc,Pawlowski:2020qer,Saueressig:2023irs} for reviews. Understanding of the resulting phenomenology is developing, see \cite{Eichhorn:2022jqj,Eichhorn:2022gku,Eichhorn:2023xee}. 
A remarkable property of such a fixed point is its near-perturbative nature, see \cite{Eichhorn:2018ydy,Falls:2013bv,Falls:2018ylp} and Fig.~15 of \cite{Eichhorn:2020sbo}.
Although its inspection typically proceeds using tools that go beyond perturbation theory\footnote{See \cite{Niedermaier:2009zz,Falls:2024noj,Kluth:2024lar} for evidence for the asymptotically safe fixed point from perturbative calculations.}, many of its properties do not greatly depart from perturbative behavior. This is advantageous for practical calculations.

Understanding neutrino masses in asymptotically safe gravity is part of a broader effort to understand fermion masses and their generation. It was found that quantum gravity does not break chiral symmetry  \cite{Eichhorn:2011pc,Meibohm:2016mkp,Eichhorn:2017eht, deBrito:2020dta, deBrito:2023kow} and can avoid gravitational catalysis \cite{Gies:2018jnv, Gies:2021upb}; thus fermions can remain light compared to the Planck mass. Based on this result, explicit mass terms were investigated \cite{Eichhorn:2016vvy, DeBrito:2019rrh} and \cite{Daas:2020dyo,Daas:2021abx} and generically found to be relevant perturbations of a fixed point, leaving the low-energy value of the mass as a free parameter.

Neutrino Yukawa couplings $y_{\nu}$ were for the first time investigated in \cite{Held:2019vmi}, see also \cite{Kowalska:2022ypk,Eichhorn:2022vgp}. 
It was found that quantum-gravity fluctuations can dynamically keep neutrino Yukawa couplings at tiny values, e.g., around $y_{\nu} \sim 10^{-10}$ or below, such that asymptotic safety may be well-compatible with very small neutrino masses.
Extended neutrino sectors were studied in \cite{Domenech:2020yjf, Chikkaballi:2023cce}, where a Seesaw scale was included for the first time. 

Against this background, we aim at answering three questions in the present paper. First, are new matter fields required to explain neutrino masses in an asymptotically safe extension of the SM by gravity?
Second, is there a theoretical bound on the Seesaw scale from quantum gravity? Third, can neutrinos be pseudo-Dirac fermions in asymptotic safety, or do they have to be either Dirac or Majorana neutrinos?

\section{Methodology}
In this work, we utilize the Wetterich equation~\cite{Wetterich:1992yh, Morris:1993qb,Ellwanger:1993mw}, reviewed in \cite{Dupuis:2020fhh} to obtain the beta functions $\beta_{g_i}= k \partial_k g_i(k)$
for the dimensionless couplings in the system. The momentum-scale $k$ stands for an infrared cutoff. It implements Wilsonian RG ideas in a smooth fashion within the modern functional framework. 
We are interested in fixed points ${g}_{i,*}$ for which $\beta_i(\{{g}_*\})=0$ for all $i$. Once such a fixed point is found, we linearize the flow around it to define the critical exponents as the eigenvalues of $-\p \beta_i(\{{g}\})/\p{g}_j|_{{g}_i={g}_{i,*}}$.
A coupling with a positive $\theta_i$ is relevant. Once the flow departs from the scale-invariant fixed-point regime, the value of a relevant coupling is a free parameter.
In contrast, an irrelevant coupling has a negative $\theta_i$. 
Such a coupling is calculable at all scales and only acquires a dependence on the scale $k$ through its dependence on the relevant parameters. Its infrared value is a prediction of the theory, making an asymptotically safe theory falsifiable, even if all available experimental data pertains to energy scales which are much lower than the scale at which (approximate) fixed-point scaling sets in.

Thereby, some free parameters of the SM (and BSM systems) become calculable from first principles in asymptotically safe gravity-matter models. An example is the Higgs quartic coupling, which sets the ratio of the Higgs mass to the electroweak scale~\cite{Shaposhnikov:2009pv}. Other free parameters of the SM are bounded from above, because quantum fluctuations of gravity and matter compete and generate two different RG fixed points, a free and an interacting one. As a consequence, couplings cannot exceed upper bounds; examples include the top and bottom Yukawa coupling~\cite{Eichhorn:2017eht, Eichhorn:2018whv} and the hypercharge coupling~\cite{Eichhorn:2017lry}, see also the reviews~\cite{Eichhorn:2022gku,Eichhorn:2023xee}.
We now explore the predictive power of asymptotic safety for neutrino mass-generation mechanisms.

\section{The necessity of degrees of freedom beyond the Standard Model in asymptotic safety}
There is an interaction which endows neutrinos with mass without introducing degrees of freedom beyond the SM. This is the Weinberg operator \cite{Weinberg:1979sa,Weinberg:1980bf}, a dimension-five operator built purely out of the left-handed neutrinos of the SM and the Higgs field $H$,
\begin{align}
\mathcal{L}_{\rm Weinberg} = \frac{\zeta}{k} \,\Big( (\bar{L} \sigma_2 H^*) (H^\dagger \sigma_2 L^C)  + \text{h.c.}\Big) \,,
\label{eq:Weinberg}
\end{align}
where $L$ are the left-handed lepton doublets, $C$ denotes charge conjugation and $\sigma_2$ is a Pauli matrix. $\zeta$ denotes a dimensionless coupling and $k$ is an energy scale introduced because the Weinberg operator is a dimension-five operator.
In the framework of the Wetterich equation, this scale is identified with the RG scale.
As a dimension-five operator, the Weinberg operator is perturbatively non-renormalizable and therefore usually not considered as part of a fundamental theory.
Instead, it is usually viewed as part of the SM Effective Field Theory (SMEFT) in which higher-order operators are generated by heavy degrees of freedom that are integrated out~\cite{Buchmuller:1985jz,Grzadkowski:2010es}. 
We take a rather different perspective, motivated by the fact that asymptotic safety changes the power-counting in a theory. In an asymptotically safe theory, perturbative nonrenormalizability is a mirage, and dimension-five (or higher) operators can be renormalizable, i.e., may become relevant perturbations of an interacting fixed point~\cite{Brenner:2024bps}.
Even if they are not relevant, they can be included into the theory, because the high-energy behavior of the corresponding coupling is determined by the fixed point, whereas the low-energy value is a prediction.
This opens the possibility that the Weinberg operator is present already at the Planck scale in the theory with only SM and gravitational degrees of freedom. 

One interpretation of this scenario is to view gravitons (despite their masslessness) as the ``heavy" degrees of freedom, because they come with an effective cutoff scale, the Planck scale. Upon integrating out gravitons, dimension-five operators are generated. We explore whether the Weinberg operator is one of them.
If this is the case, a remarkable conclusion could be drawn: neutrino masses and neutrino oscillations could be accommodated in the SM with gravity without new matter fields. This would surely be the most ``economical" solution to the neutrino-mass problem. Therefore it is important to understand whether or not it is realized, before extra degrees of freedom -- so far not seen by experiments -- are postulated to exist.

We parameterize gravitational fluctuations through the dimensionless Newton coupling $G$ which assumes a nonzero constant value at trans-Planckian scales, see the  
appendix
for details. 
The resulting beta function reads
\begin{eqnarray}
\beta_{\zeta}&=&\zeta - \frac{3}{16\pi^2}g_2^2\, \zeta + \frac{3}{8\pi^2}\left( y_t^2 + y_b^2 - \frac{1}{6}y_{\tau}^2\right) \zeta \nonumber\\
&{}&+ \frac{1}{4 \pi^2}\lambda_H\, \zeta + \frac{17}{18\pi}G\, \zeta.\label{eq:beta_zeta}
\end{eqnarray}
The non-gravitational part, neglecting the tiny contributions from the first two generations, agrees with \cite{Zhang:2024weq}; the gravitational part (in the last term) is a new result. The first term encodes how $\zeta$ scales to zero canonically, because the Weinberg-operator is a dimension-five-operator and thus highly suppressed in the absence of other interactions. The SM interactions cannot counteract this effect, because the screening contribution of $y_{t/b}$ and $\lambda_H$ overwhelms the antiscreening part and because both are numerically small compared to the first term.
We discover that gravity fluctuations also screen the coupling.
Thus, any non-zero initial value for $\zeta$ is driven towards zero.

In an asymptotically safe setting, the initial value $\zeta(k \gtrsim M_{\rm Pl})$ is given by the value at an RG fixed point, $\zeta_{\ast}$.
As a consequence of Eq.~\eqref{eq:beta_zeta} being linear in $\zeta$, there is only the fixed point $\zeta_{\ast}=0$, at which $\zeta$ is irrelevant\footnote{The near-perturbative nature of the fixed-point for gravity-matter systems is crucial to neglect the contribution of additional, higher-order interactions in our considerations.}, with critical exponent $\theta_{\zeta} \approx -1+\frac{17}{18\pi G}$ (if all other SM couplings vanish at the fixed point.)
Hence it cannot depart from the fixed-point value as long as gravity fluctuations are present.
Thus, the only value compatible with an asymptotically safe regime above the Planck scale is $\zeta(k = M_{\rm Pl})=0$. In the appendix we investigate the robustness of this result and find that it is robust under variations of the gravitational fixed-point values.
At a near-perturbative fixed point, as the gravity-matter fixed point appears to be, quantum fluctuations are not strong enough to turn the dimension-five Weinberg operator into a relevant operator and thus the Planck-scale value of its coupling must vanish.

Because this value results in lepton-number conservation of the SM (in the absence of BSM degrees of freedom, most importantly right-handed neutrinos), it is a protected value and thus $\zeta(k)=0$ holds also for all $k< M_{\rm Pl}$. 
This leads us to predict that the asymptotically safe SM with gravity has exactly vanishing Weinberg operator at all scales. To obtain nonzero neutrino masses in an asymptotically safe setting, \emph{there must be extra degrees of freedom beyond gravity and the SM fields}. 
To the best of our knowledge, this is the first unequivocal evidence for such degrees of freedom in asymptotic safety.

\section{Can neutrino masses arise from a type-I Seesaw mechanism in asymptotic safety?}
We have just shown that an asymptotically safe theory \emph{must} contain new degrees of freedom beyond the SM to account for neutrino masses. The arguably most popular way to endow neutrinos with masses is through the Seesaw mechanisms~\cite{Minkowski:1977sc,Yanagida:1979as,Yanagida:1980xy}.
The basis for the type-I Seesaw mechanism is the assumption that there are right-handed neutrinos $\nu_R^i$ and the neutrino Yukawa couplings should not be much smaller than the other Yukawa couplings in the SM.
 This can be achieved by adding a Majorana mass term for the right-handed neutrinos, such that the neutrino mass matrix becomes
\begin{eqnarray}
\mathbb{M} = 
\begin{pmatrix}
0 & m_D\\
m_D & m_R
\end{pmatrix},
\label{eq:mass matrix}
\end{eqnarray}
where $m_R$ is the mass parameter introduced for the Majorana mass term and $m_D$ is a mass parameter associated with the Dirac mass term. The eigenvalues of the mass matrix are
\be
m_{1,2}=\frac{m_R}{2}\pm \frac{1}{2}\sqrt{m_R^2+4 m_D^2}\,.
\label{eq:mass eigenvalues}
\ee
For $m_R \gg m_D$, we obtain $|m_1| \approx m_R$ and $|m_2| \approx m_D^2/m_R$.
Thus, the eigenvalue $m_1$ accounts for the mass of a heavy (still undetected) neutrino, while the eigenvalue $m_2$ accounts for the mass of one of the light neutrinos existing in our Universe. 
Both eigenstates of the mass matrix are Majorana neutrinos. 
Therefore, neutrinoless double beta decay is allowed \cite{Doi:1980yb,Langacker:1986jv,Giunti:2010ec}, see \cite{Klapdor-Kleingrothaus:2001oba,Phillips:2011db,NEMO-3:2015jgm,NEXT:2015wlq,KamLAND-Zen:2016pfg,EXO:2017poz,CUORE:2019yfd,EXO-200:2019rkq,GERDA:2020xhi} for experiments that search for it.

The SHiP experiment, to be constructed at CERN, will probe part of the parameter space for the right-handed heavy neutrinos, for mass-scales close to the TeV-range~\cite{SHiP:2015vad}. 
If the Yukawa coupling is $\mathcal{O}(1)$, one is naturally led to a mass scale $m_R \sim 10^{16}\, \rm GeV$. Given the 13 orders of magnitude between the TeV scale at $10^{16}\, \rm GeV$, additional considerations that can further limit the viable range of masses are highly desirable.

A lower bound arises, if one makes the assumption that the SM supplemented by right-handed neutrinos should also explain dark matter and the matter-antimatter asymmetry.
The heavy mass eigenstate is a sterile, right-handed heavy neutrino, and an, either completely stable~\cite{Boyle:2018rgh} or long-lived~\cite{Asaka:2005pn} dark-matter candidate~\cite{Boyarsky:2018tvu}. In such a dark-matter scenario, there is a lower mass-bound, the Lee-Weinberg bound~\cite{Lee:1977ua,Sato:1977ye}, which reads $m_R>2$~GeV. In addition, the matter-antimatter-asymmetry in the universe can be explained through leptogenesis~\cite{Fukugita:1986hr}; see also \cite{Buchmuller:1999cu,Davidson:2008bu} and references therein. 
The thermal leptogenesis scenario establishes a lower mass-bound on the lightest right-handed neutrino mass, known as the Davidson-Ibarra bound~\cite{Davidson:2002qv}, at $m_R>10^{8-9}$~GeV.  An upper bound stems from the reheating temperature such that $m_R<T_\text{reh}$. 
Unfortunately, the reheating temperature is not known and only bounded from below: numerical and analytic studies~\cite{Buchmuller:1999cu,Josse-Michaux:2007alz,Davidson:2008bu} lead to $T_\text{reh}>10^{10}$~GeV such that no quantitative upper bound follows from these considerations.

We now derive a -- to the best of our knowledge first -- quantitative upper bound on $m_R$. We do so by requiring a consistent coupling to quantum gravity. It holds, irrespective of whether or not the right-handed neutrino is a dark-matter candidate and irrespective of whether a thermal leptogenesis scenario is assumed.

The Lagrangian for the neutrino sector in the type-I Seesaw model is given by 
\begin{equation}
    \mathcal L_\nu = y_{\nu} \bar\ell H \nu_R+ \frac{m_R}{2} \bar \nu_R^C \nu_R  + {\rm h.c.}
\end{equation}
where $(\nu_R)^C$ is the charge conjugated right-handed neutrino field, $H$ is the Higgs field and $y_{\nu}$ is the Yukawa coupling. To exhibit the mechanism that gives rise to an upper bound on the Seesaw scale, we limit ourselves to one generation of SM fields and neglect mixing of the neutrinos. After electroweak symmetry breaking, the Dirac-Yukawa coupling gives rise to the Dirac mass $m_D=y_{\nu} v_H/\sqrt{2}$ with the Higgs vacuum expectation value $v_H$.

The beta functions that describe the running couplings of the third generation of fermions in the SM, together with the gravitational contribution and the Majorana mass, are presented in an ancillary notebook. For $m_R \rightarrow 0$, these are in agreement with the beta functions in \cite{Eichhorn:2018whv} and \cite{Eichhorn:2022vgp}.

We confirm the results from \cite{Eichhorn:2022vgp} and extend them to nonvanishing $m_R$. In particular, we discover that there is an asymptotically free fixed point in all Yukawa couplings at which $m_R=0$ holds. At this fixed point, $m_R$ is also asymptotically free, i.e., it can increase towards the infrared and its low-energy value is a free parameter. 
Thus, from this fixed point, the experimental values of the Yukawa couplings can be reached at low energies and $m_R$ increases. Its growth is, however, limited, due to an upper bound on all Yukawa couplings that arises in asymptotic safety, see \cite{Eichhorn:2017eht, Eichhorn:2018whv}. This bound is given by an \emph{infrared attractive} fixed point in any given Yukawa coupling. This fixed point gives rise to an upper bound on the Yukawa coupling at the Planck scale: any value beyond the upper bound is shielded from the asymptotically free fixed point and cannot correspond to a UV complete theory \cite{Eichhorn:2017ylw}. To illustrate this, let us consider the beta function for the neutrino Yukawa coupling on its own, while neglecting all other couplings in the SM. It reads
\begin{equation}
\beta_{y_{\nu}}=-f_y \, y_{\nu} + \frac{5}{32\pi^2}y_{\nu}^3, 
\end{equation}
with
\begin{align}
f_y &= -G\frac{96 -235\Lambda +103\Lambda^2+56\Lambda^3}{12 \pi (3-10\Lambda +  8\Lambda^2)^2}.
\end{align}
Its value at the fixed point is 
\begin{align}
f_y|_{G \rightarrow G_{\ast} = 4.6 ~ \Lambda \rightarrow \Lambda_{\ast}=-6.8} = 7 \cdot 10^{-3},
\end{align}
see \cite{Eichhorn:2017eht}, which is very small, as one would expect from a near-perturbative fixed point.\footnote{The couplings $G$ and $\Lambda$ are dimensionless counterparts of the Newton and cosmological constants where the (functional) RG scale $k$ is used as the conversion unit between dimensionful and dimensionless couplings; $G_{\ast}$ and $\Lambda_{\ast}$ denote their fixed-point values.} In turn, this limits $y_{\nu}(k= M_{\rm Planck})\lesssim 0.66$. The RG flow between the Planck scale and the electroweak scale translates this into a limit at the electroweak scale.

Here, we discover that there is a consequence for the right-handed neutrino mass and the corresponding eigenvalue $m_1$ of the mass matrix, namely an upper bound on $m_R= m_D^2/m_2$. 
The eigenvalue $m_2$ is bounded from above by observations. In asymptotically safe gravity,  $m_D$ is bounded from above, because $y_{\nu}$ is bounded from above.\footnote{In \cite{Chikkaballi:2023cce}, it was noted that a fixed point, at which $y_{\nu}$ is non-zero and irrelevant (i.e., corresponds to a prediction of the theory), fixes $m_R$ in terms of $m_2$. Here we note for the first time, that there is  a \emph{global} consequence for the RG flow, namely trajectories that start out from the free fixed point are subject to the upper bound on the right-handed mass.}
We therefore obtain
\be
m_R \lesssim \frac{y_{\nu, \, \rm upper}^2 \, v_H^2}{2 m_{2}},
\label{eq:upperbound}
\ee
where $m_2$ is the mass of the light neutrino and $y_{\nu, \, \rm upper}$ is the upper bound from asymptotic safety. To provide concrete numbers, we make the assumption that $m_{2}$ takes a definite value; in practice there is of course only an upper bound on the mass available.
For concreteness, we use gravitational fixed-point values $G_{\ast}=4.6$ and $\Lambda_{\ast}=-6.8$, as they arise from the beta functions in \cite{Eichhorn:2017ylw} with a matter content corresponding to the SM plus three right-handed neutrinos. We then find that $y_{\nu}(k=m_t=173\, {\rm GeV})< 0.45$ has to hold, if all the other SM couplings flow from an asymptotically free fixed point to their observed values in the IR. Using a neutrino mass $m_{2}= 10^{-10}\, \rm GeV$ as an example, we find the bound
\be
m_R< \frac{(0.45\cdot 246\, {\rm GeV})^2}{2 m_{2}} \simeq \frac{1.2\cdot 10^{4} \,{\rm GeV}^2}{2\cdot 10^{-10} \,\rm GeV}
= 6\cdot 10^{13}\, \rm GeV. 
\label{eq:upper bound}
\ee
This is five orders of magnitude below the Planck scale, which a priori constitutes the upper limit on particle masses in an asymptotically safe quantum theory of gravity. It is also two orders of magnitude below the GUT scale; a scale that is often associated with right-handed neutrino masses~\cite{Albright:2001xq,Albright:2001pj}.

\begin{figure}[!t]
\includegraphics[width=\linewidth]{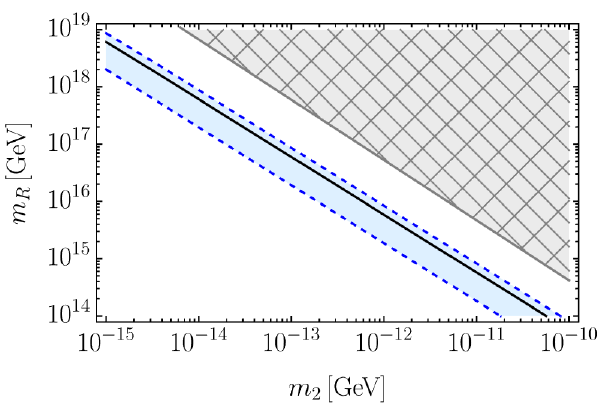}
\caption{\label{fig:upperBoundband}
We show the upper bound on $m_R$ as a function of $m_2$. The solid (black) line indicates the upper bound obtained at the gravitational fixed point where $G_* = 4.6$ and $\Lambda_* = -6.8$. The blue strip between the dashed lines corresponds to variation of the
fixed point $G_*$ by $33.3 \%$ above and below its computed value. 
The upper-right (meshed) region is excluded, because it cannot be reached from a UV complete theory with gravity; instead,
the RG trajectories exhibit a Landau pole below the Planck scale.
}
\end{figure}

The quantitative value of this bound is subject to systematic uncertainties, mainly through the systematic uncertainty in the gravitational fixed-point values, which translates into a corresponding uncertainty of the upper bound on $y_{\nu}$.
We bound this systematic uncertainty from below by using results from \cite{Dona:2013qba}. There, the gravitational fixed-point value varies by about $1/3$ of its absolute value, when the regulator function is changed. We show the upper bound on $m_R$ as a function of $m_2$ within the resulting band in Fig.~\ref{fig:upperBoundband}.

\begin{figure*}[!t]
\includegraphics[width=\linewidth]{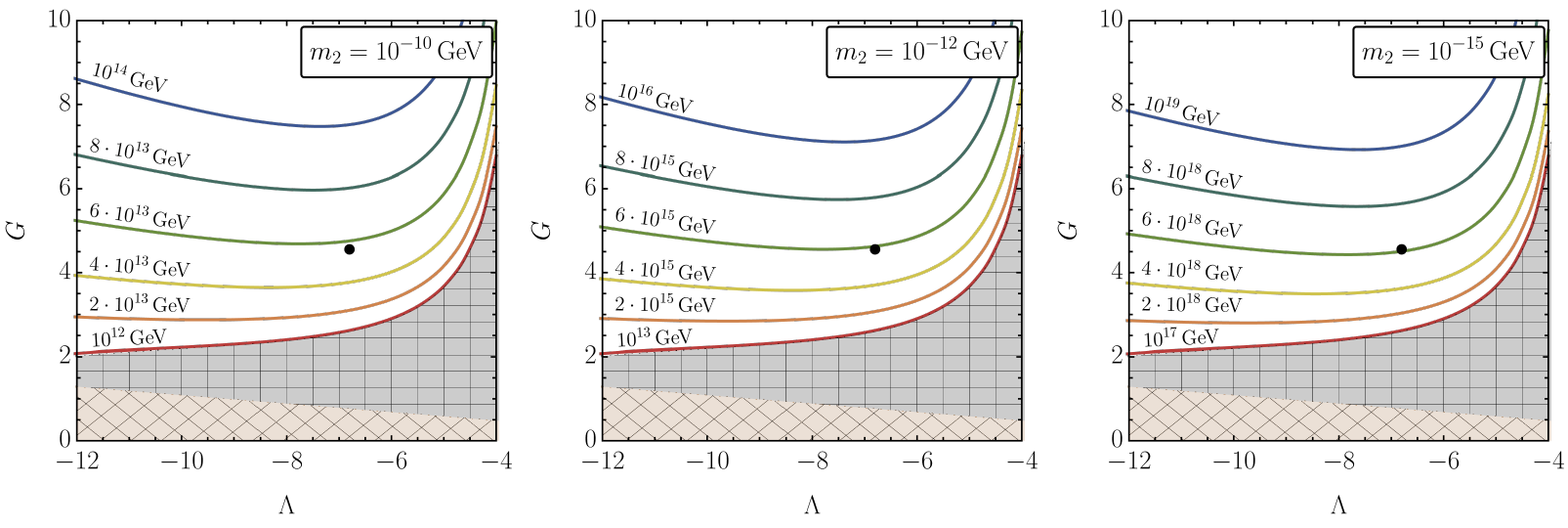}
\caption{\label{fig:upperBound}
We show the dependence of the upper bound for $m_R$ from Eq.~\eqref{eq:upperbound} on $G$ and $\Lambda$.
The central black dot indicates the fixed-point value $G_*=4.6$ and $\Lambda_*=-6.8$, for which the upper bound becomes Eq.~\eqref{eq:upper bound}. The quantitative value of the upper bound Eq.~\eqref{eq:upperbound} also depends on the observed, light neutrino mass. We show three different values for the mass, $m_2 = 10^{-10}\, \rm GeV$, $m_2 = 10^{-12}\, \rm GeV$ and $m_2 = 10^{-15}\, \rm GeV$. 
The gray region arises from the requirement that the Abelian gauge coupling and/or the top Yukawa coupling in the SM can be accommodated; in the gray region, the upper bounds on one or both of these couplings are below the observed values.
}
\end{figure*}

The existence of this bound is, however, robust for a viable theory including the SM.\footnote{The upper bound arises due to an 
antiscreening contribution that arises due to gravitational effects in the beta function of Yukawa couplings that generates a relevant perturbation of the fixed point associated with the Yukawa coupling. In our truncation, it is encoded in $f_y>0$, but in extended truncations, the physics generating the relevant direction may be encoded in other ways \cite{debritotoappear}.}
The dependence of the upper bound~\eqref{eq:upperbound} on $G$ and $\Lambda$ is displayed in Fig.~\ref{fig:upperBound}. 
The mass-range required for thermal leptogenesis can be accommodated everywhere in the parameter space that is not already ruled out by other considerations.

A right-handed neutrino mass several orders of magnitude below the Planck scale is compatible with asymptotic safety, but requires a value of $m_R (k=M_{\rm Planck})/M_{\rm Planck} \approx 10^{-5}$. To such a tiny value, one can have one of two attitudes: either, one regards any value for a free parameter as equally viable, or one requires that dimensionless ratios should be $\mathcal{O}(1)$ in the absence of a dynamical mechanism to restrict their value. The latter attitude amounts to a requirement that there be no ``fine-tuning" of RG trajectories\footnote{Strictly speaking, such statements require a measure on the space of RG trajectories, to determine whether any given RG trajectory is more or less ``fine-tuned" than any other.}.
In this spirit, if one were to require a value $m_R/M_{\rm Planck} \approx 1$, the inequality \eqref{eq:upperbound} can be turned into a prediction for the observed neutrino mass, namely $m_{2}\lesssim 2 \cdot 10^{-15}\,\rm GeV$.

\begin{figure}[!t]
\includegraphics[width=\linewidth]{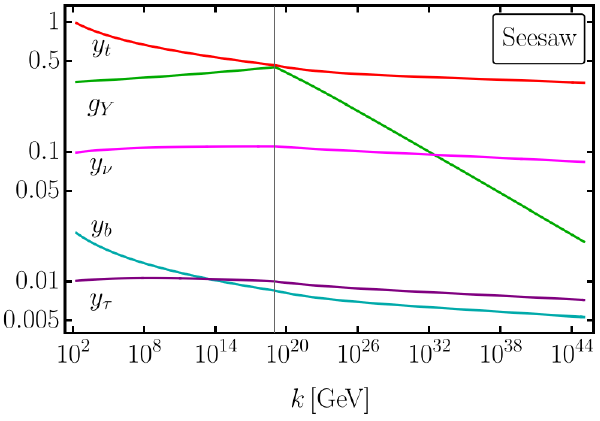}
\caption{\label{fig:runningcouplings} We show the RG flow of the SM couplings and the neutrino Dirac-Yukawa coupling from trans-Planckian to sub-Planckian scales in the type-I Seesaw scenario. The Planck-scale, at which gravitational fluctuations switch off dynamically, is indicated by a thin black line. 
In the very far UV, not shown in this figure, all couplings emanate from an asymptotically free fixed point.}
\end{figure}

To provide a concrete example of UV-complete trajectories within this setup, we can realize $m_{2}= 10^{-10}\, \rm GeV$ with $y_{\nu}(k=173\,{\rm GeV})=0.1$, $m_R(k=173\,{\rm GeV})=2.9 \cdot 10^{12}\, \rm GeV$ 
and the other SM couplings fixed to their one-loop values at $k=173\, \rm GeV$, namely $g_Y=0.35$, $g_2=0.65$, $g_3=1.17$, $y_t=0.996$, $y_b=0.024$, $y_{\tau}=0.01$. The corresponding RG flow, emanating from the asymptotically free fixed point in the UV (with the gravitational couplings at their asymptotically safe fixed-point values given above), is shown in Fig.~\ref{fig:runningcouplings}.

\section{Can neutrinos be Pseudo-Dirac neutrinos in asymptotic safety?}
There is another option for neutrino masses, namely that the Majorana mass of the right-handed neutrinos is very small compared to the Dirac mass~\cite{Wolfenstein:1981kw,Petcov:1982ya,Bilenky:1983wt}. 
In such a case, the eigenvalues~\eqref{eq:mass eigenvalues} of the mass matrix~\eqref{eq:mass matrix} are such that $|m_{1,2}| \approx m_D \pm m_R/2$ and the neutrino essentially becomes a Dirac particle. 
The right-handed Majorana mass resolves the degeneracy of the mass eigenstates only very slightly and correspondingly, lepton-number violation is a very small effect, leading to a highly suppressed neutrinoless double-beta decay. The main experimental pathway to search for pseudo-Dirac is through oscillations between active and sterile neutrinos driven by the tiny mass splitting between the mass eigenstates. See \cite{Kobayashi:2000md} for the phenomenological implications and the possible theoretical framework for the Pseudo-Dirac-scenario. 
Experimental searches for Pseudo-Dirac neutrinos are possible at JUNO~\cite{Franklin:2023diy} and DARWIN~\cite{deGouvea:2021ymm}.

\begin{figure}[!t]
\includegraphics[width=\linewidth]{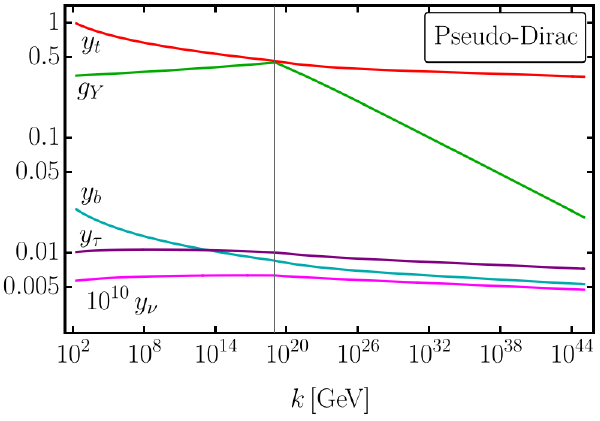}
\caption{\label{fig:PD} We show an example for the RG flow of the SM couplings and the neutrino Dirac-Yukawa coupling from trans-Planckian to sub-Planckian scales in the Pseudo-Dirac neutrino scenario. The Planck-scale, at which gravitational fluctuations switch off dynamically, is indicated by a thin black line. The neutrino Yukawa coupling is rescaled by a factor of $10^{10}$ to make it of a similar order of magnitude as the other SM couplings.}
\end{figure}

In asymptotically safe gravity, the Majorana mass term is a relevant perturbation of the fixed point. Thus we expect that we can freely choose its low-energy value (below its upper bound), such that a Pseudo-Dirac neutrino can be realized, cf.~\cite{Chikkaballi:2023cce}. We test this by explicitly constructing RG trajectories for which $m_R = 10^{-2} m_D$, cf.~Fig.~\ref{fig:PD}. 
We stress that there are two free parameters that determine the neutrino masses in this spectrum: first, the neutrino Yukawa coupling is freely adjustable, as long as its IR value is much lower than $\mathcal{O}(1)$, which it is in the present setting. Second, the right-handed neutrino mass is also freely adjustable and can thus be shifted anywhere in the range between 0 GeV and the Planck mass (as long as the observable mass is not fixed). It is thus not unexpected that we can accommodate a spectrum with a tiny pseudo-Dirac mass for neutrinos. 

We thus find that asymptotically safe gravity is compatible with a Pseudo-Dirac scenario. We do not find any constraints on the parameters of the neutrino sector of the theory in this setting.
In other words, Pseudo-Dirac neutrinos lie in the ``landscape" of asymptotic safety. For other theories of quantum gravity this may be similar; in fact, there is a general argument that quantum gravity generates tiny right-handed neutrino masses through higher-order, Planck-scale suppressed operators, see, e.g., \cite{Babu:2022ikf}. Because these are general arguments, not based on derivations from particular quantum theories of gravity, we do not yet know enough about whether Pseudo-Dirac-neutrinos lie in the swampland of any approach to quantum gravity or whether they are part of a universal landscape, see \cite{Eichhorn:2024rkc}.

\section{Conclusions}
In this work, we have tested a set of models of neutrino mass generation for their compatibility with asymptotically safe gravity. In case of compatibility, we have explored whether there are theoretical constraints on the parameters of the models.
As an overarching result, we find that constraints exist, and some models are even ruled out. This may at a first glance be surprising, due to the huge separation between the mass-scale of light neutrinos and the Planck scale. It is, however, a consequence of two insights. The first insight is that asymptotic safety has predictive power, even for couplings which are marginal without gravity and thus free parameters of the SM and its extensions. This predictive power manifests in predictions and upper bounds on couplings at the Planck scale. The second insight is that the RG is a tool that bridges large gaps in scales, including the gap between the Planck scale and the mass scale of light neutrinos. In practice, the RG can thus be used to translate predictions on the couplings at the Planck scale into resulting predictions at experimentally accessible scales.

Our second key result is that neutrinos must remain massless, unless new degrees of freedom are added to the SM with gravity. This is because the Weinberg operator corresponds to an irrelevant perturbation of the fixed point, at which the corresponding coupling vanishes. This entails that the coupling vanishes at all scales. Theories in which neutrino masses arise from the Weinberg operator at a fundamental level are therefore in the swampland of asymptotically safe quantum gravity.\footnote{This does not rule out that the Weinberg operator arises in the EFT by integrating out non-gravitational new physics below the Planck scale.} 
This implies that there must be new degrees of freedom beyond the SM in any theory that describes nature (and thus neutrino oscillations) and that is asymptotically safe. 
To the best of our knowledge, this is the first robust evidence that the SM together with gravity is insufficient to describe observations.\footnote{ Evidence for dark matter and the observed matter-antimatter asymmetry must of course also be explained in such a theory. It is, however, not robustly excluded that this may proceed without extra fields, e.g., through primordial black holes as dark matter and an asymmetry between matter and antimatter that arises as an initial condition in the early universe after a quantum-gravity dominated phase.}

Our third key result is that there is an upper bound on the mass scale of right-handed neutrinos, such that type-I Seesaw models with a sufficiently high mass scale are in the swampland.
This is, to the best of our knowledge, the first time that an upper bound on the mass scale of right-handed neutrinos has been found.

Finally, we find that two models are in the asymptotically safe landscape, namely type-I Seesaw models with a low enough Seesaw scale as well as models in which neutrinos are of Pseudo-Dirac type.

Taking a broader perspective, our results imply that experiments that probe the nature of and dynamical mechanism underlying neutrino masses also indirectly shed light onto the quantum structure of spacetime. This perspective is supported by results that constrain neutrino masses from string-theory-inspired swampland conjectures~\cite{Gonzalo:2021fma,Gonzalo:2021zsp,Harada:2022ijm}, results on neutrinos mass from string theory, see \cite{Casas:2024clw} and references therein and by previous results on neutrino Yukawa couplings in asymptotically safe gravity~\cite{Held:2019vmi,Eichhorn:2022vgp,Kowalska:2022ypk}.

\begin{acknowledgements}
G.\,P.\, B.~and M.\,Y. acknowledge the Institute for Theoretical Physics, Heidelberg University, for the very kind hospitality during their stay.
The work of G.\,P.\, B.~and A.~E.~was supported by
the research grant (29405) from VILLUM fonden during most of this project.
The work of A.\,D.\, P. is supported by CNPq under the grant PQ-2 (312211/2022-8) and FAPERJ under the ``Jovem Cientista do Nosso Estado'' program (E-26/205.924/2022).
The work of M.\,Y. was supported by the National Science Foundation of China (NSFC) under Grant No.~12205116, the Seeds Funding of Jilin University, and the Humboldt Alumni Program.
\end{acknowledgements}
\bibliography{refs}

\clearpage
\onecolumngrid
\appendix
\section{APPENDIX}

\subsection{Choice of regulator in the functional Renormalization Group}
In this work, we employ the Litim-type cutoff function~\cite{Litim:2001up}
\begin{align}
\mathcal R_k(p) = 
\begin{cases}
\displaystyle 
p^2\left( \frac{k^2}{p^2}-1\right) \theta(p^2-k^2) & \text{for bosons}\,,\\[4ex]
\displaystyle
\Slash p\left( \sqrt{\frac{k^2}{p^2}} -1\right) \theta(p^2-k^2) & \text{for fermions}\,.
\end{cases}
\end{align}
In what follows, we make an ansatz for the effective action $\Gamma_k$ to derive the beta functions for couplings.

\subsection{General truncation for gravity-matter systems}
We make the following ansatz for the effective action:
\begin{align}
\Gamma_k = \Gamma_k^\text{SM} + \Gamma_k^\text{grav} + \Gamma_k^\nu.
\end{align}

For the gravity sector, we assume the Einstein-Hilbert action in Euclidean signature, i.e.,
\begin{align}
\Gamma_k^\text{grav} = \frac{k^2}{16\pi G} \int \df^4x\sqrt{g} \left[ -R + 2\Lambda k^2 \right] + S^\text{grav}_\text{gh+gf},
\label{appeq:gravity sector}
\end{align}
where $G$ and $\Lambda$ are the dimensionless Newton constant and the dimensionless cosmological constant, respectively, $\sqrt{g}$ is the determinant of the metric field $g_{\mu\nu}$, and $R$ is the Ricci curvature scalar.
To calculate gravitational loop effects, we split the metric field $g_{\mu\nu}$ into a background field $\bar g_{\mu\nu}$ and a fluctuation field $h_{\mu\nu}$, i.e., 
\begin{align}
g_{\mu\nu} = \bar g_{\mu\nu} + h_{\mu\nu}.
\end{align}
Following the standard Faddeev-Popov quantization procedure, we introduce the gauge and ghost field actions denoted by $S^\text{grav}_\text{gh+gf}=S^\text{grav}_\text{gh}+S^\text{grav}_\text{gf}$. 
Here, these actions are given respectively by
\begin{align}
S^\text{grav}_\text{gh}&= -\int\df^4x\sqrt{\bar g}\,\overline C_\mu
               \left[ {\bar g}^{\mu\rho}{\bar D}^2+
               \frac{1-\beta}{2}{\bar D}^\mu{\bar D}^{\rho}
               +{\bar R}^{\mu\rho}\right] C_{\rho}\,,
               \label{ghostaction}
\\
S^\text{grav}_\text{gf}&= \frac{1}{2\alpha}\int \df^4x\sqrt{\bar g}\,
                       \left( {\bar D}^\nu h_{\nu \mu}-\frac{\beta +1}{4}{\bar D}_\mu h\right)^2 \,,
                        \label{gauge fixing action}
\end{align}
where $\bar D$ is the covariant derivative constructed only by background metric $\bar g_{\mu\nu}$, and $C_\mu$ and $\overline C_\mu$ are the ghost and anti-ghost vector fields, respectively.
We employ the harmonic gauge fixing condition for the metric fluctuations in which there are two gauge fixing parameters denoted by $\alpha$ and $\beta$, we choose $\beta = \alpha \rightarrow 0$ (Landau-DeWitt gauge). 

Next, we consider the SM action. Its effective action consists of four parts:
\begin{align}
\Gamma_k^\text{SM}= \Gamma_k^\text{gauge} + \Gamma_k^\text{fermion} + \Gamma_k^\text{Yukawa} + \Gamma_k^\text{Higgs}.
\end{align}
Here, the action for the gauge sector is given by
\begin{align}
\Gamma_k^\text{gauge}  &=  \frac{1}{4g_3^2}\int\df^4x\sqrt{g} \, g^{\mu\rho} g^{\nu\sigma}G_{\mu\nu}^aG^{a}_{\rho\sigma}
+\frac{1}{4g_2^2}\int\df^4x\sqrt{g} \, g^{\mu\rho} g^{\nu\sigma}F_{\mu\nu}^aF^{a}_{\rho\sigma} \nn
&\qquad 
+\frac{1}{4g_Y^2}\int\df^4x \sqrt{g} \, g^{\mu\rho} g^{\nu\sigma}B_{\mu\nu}B_{\rho\sigma} 
+ S^\text{gauge}_\text{gh+gf}\,,
\end{align}
in which $G_{\mu\nu}^a$, $F_{\mu\nu}^a$ and $B_{\mu\nu}$ are the field strength tensors for $SU(3)_c$, $SU(2)_L$ and $U(1)_Y$ gauge fields with gauge couplings $g_3$, $g_2$ and $g_Y$, respectively, and the corresponding gauge fixing terms and ghost field actions are denoted by $S^\text{gauge}_\text{gh+gf}$. 
In this work, we employ the Landau gauge condition for all the gauge fields.

The Higgs sector is given by
\begin{align}
\Gamma_k^\text{Higgs} = \int \df^4x\sqrt{g}\left[
g^{\mu\nu} (D_\mu H)^\dagger (D_\nu H) + m_H^2 k^2 (H^\dagger H) + \lambda_H(H^\dagger H) ^2
\right],
\end{align}
 The covariant derivative contains the standard coupling to the electroweak gauge bosons.
 $m_H^2$ is the dimensionless Higgs-mass parameter and the Higgs field is parametrized as
\begin{align}
H&=(H)^a= \pmat{H^1\\H^2} 
= \frac{1}{\sqrt{2}}\pmat{\chi_1+i\chi_2\\[1ex] h+ i\chi_3}.
\label{eq:Higgs field}
\end{align}

For the fermion part, we take into account the top ($t$) and bottom ($b$) quark fields, and the tau-neutrino ($\nu_\tau$) and tauon ($\tau$) fields whose kinetic terms are summarized in $\Gamma_k^\text{fermion}$ as
\begin{align}
\Gamma_k^\text{fermion}= \int \df^4x\sqrt{g} \,\, \bar\psi i\Slash \nabla \psi  ,
\end{align}
where $\Slash \nabla= \gamma^\mu \nabla_\mu$ is the covariant derivative acting on a Dirac fermion $\psi$ in the appropriate representation of the SM gauge group. We also include the spin connection which encodes the minimal coupling to gravity. We translate fluctuations of the corresponding vielbein field to fluctuations of the metric, see, e.g., \cite{Eichhorn:2011pc}.

The Yukawa interactions with the Higgs field are given in $\Gamma_k^\text{Yukawa}$ as
\begin{align}
\Gamma_k^\text{Yukawa} = \int\df^4x\sqrt{g} \left[ y_t\bar Q \widetilde H t_R + y_b\bar Q H b_R + y_\tau \bar L H \tau_R + \text{h.c.}
\right],
\end{align}
with the Yukawa couplings $y_t$, $y_b$ and $y_\tau$.
Here, $\widetilde H = i\sigma^2 H^*$, and the left-handed quark and lepton double fields are
\begin{align}
\label{appeq:doublet fields}
&Q = (Q_\alpha)^a
= \pmat{ Q^{1} \\ Q^{2}}_\alpha 
=\pmat{t_{L} \\[1ex] b_L},&
&L= (L_\alpha)^a
= \pmat{ L^{1} \\ L^{2}}_\alpha 
=\pmat{\nu_{L}\\[1ex] \tau_L}_\alpha\,,
\end{align}
with $SU(2)_L$ indices $a$, $b$, $c$, $d$, and two-component spinor indices $\alpha$, $\beta$. 

The neutrino sector except for the kinetic term of the neutrino field is given in
\begin{align}
\label{appeq:neutrino sector}
\Gamma_k^\nu = \int \df^4x \sqrt{g} \, \mathcal L_\nu \,.
\end{align}
This sector depends on the scenarios for neutrino masses and is specified in the main text.

\subsection{Study of systematic uncertainties}
By invoking truncations in the solution of the RG equations, we generate systematic uncertainties in our results. It is crucial to assess these to test the robustness of our conclusions. One possibility of testing the robustness is to test the response of the results to unphysical parameters, e.g., parameters in the choice of the regulator function or gauge parameters. These unphysical parameters drop out at the level of physical observables, when the RG flow is solved exactly. Thus, the variation of physical observables under changes in the unphysical parameters provides us with an estimate of the systematic uncertainty of our result within a truncation. 

\begin{figure}
\includegraphics[width=1\linewidth]{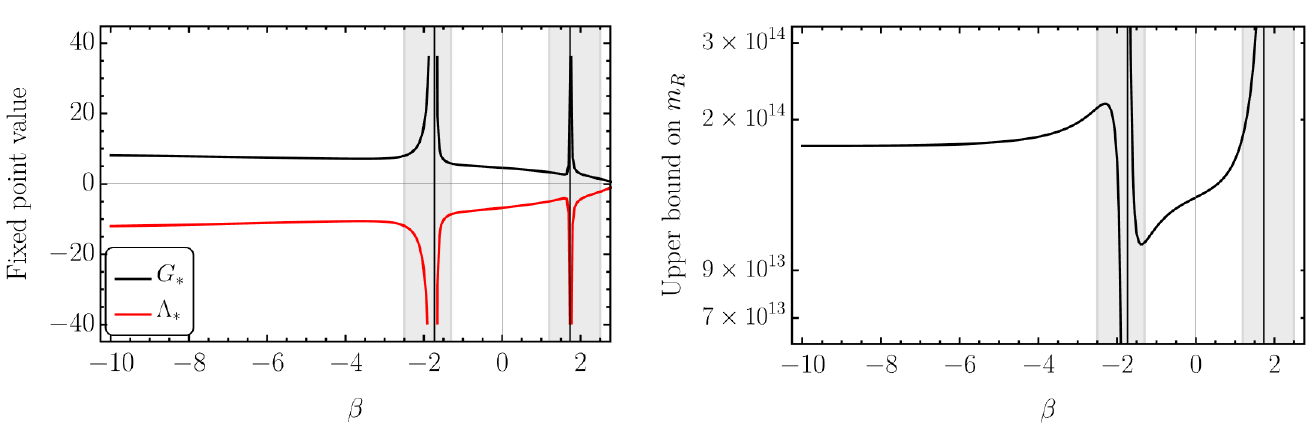}
\caption{\label{fig:FPsgauge}
Gauge-parameter ($\beta$) dependence of the fixed point value (left) and the upper bound on the right-handed Majorana mass (right) in GeV. The vertical lines indicate the poles at $\beta=\pm\sqrt{3}$.
In the right panel, we estimate the upper bound by using the fixed-point value for the neutrino Yukawa coupling as the initial condition at the Planck scale, and neglect the effects of transplanckian running.
}
\end{figure}

In this spirit, we use the dependence on the gauge parameter $\beta$ in the gravitational sector to estimate systematic uncertainties. Let us stress that the actual systematic uncertainty may also be larger than the estimate based on a single unphysical parameter. 
Before showing the gauge dependence, note that there is a pole in the propagator of a scalar mode in the metric field at
\begin{align}
\Lambda = \frac{-\beta^2+6 \beta-9}{4 \left(\beta^2-3\right)}.
\end{align}
We see from this that $\beta=\pm\sqrt{3}$ are singular points.
The gauge dependence of the fixed point value for $G$ and $\Lambda$ is shown in the left-hand side of Fig.~\ref{fig:FPsgauge}.

\begin{figure}
\includegraphics[width=.9\linewidth]{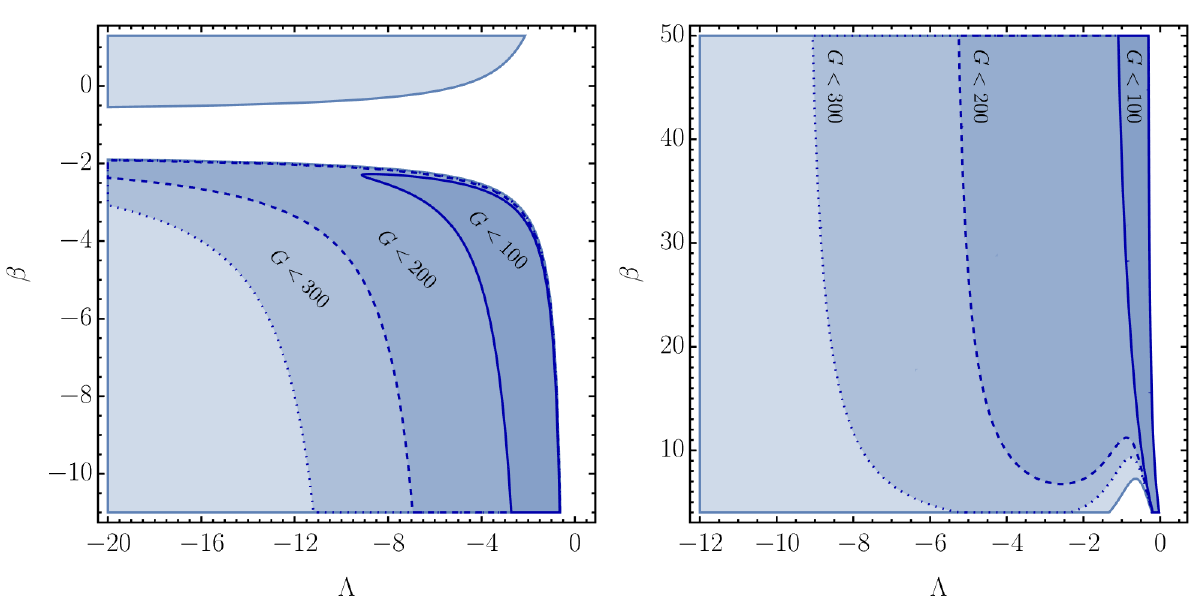}
\caption{\label{fig:Weinberggauge}We show the gravitational contribution to the critical exponent at the free fixed point in $\zeta$. The gravitational contribution is positive everywhere in the light-blue region, which spans most of the two parts of the $\beta$-$\Lambda$-plane that we show. Thus, the sign of the gravitational contribution is the correct one to make the coupling relevant. However, the gravitational contribution must overwhelm the canonical scaling term. This can only be achieved for very large values of $G$ shown in the regions surrounded by a continuous blue curve for $G<100$ and correspondingly for the other regions. We stress that in the region with $G<100$, the lowest values of G lie around $G \approx 30$ and are thus significantly higher than we expect of typical fixed-point values.}
\end{figure}

Regarding the irrelevance of the Weinberg operator, we find robustness of this result under changes of the gauge parameter, cf.~Fig.~\ref{fig:Weinberggauge}.

\end{document}